\documentstyle[12pt,twoside,fleqn,espcrc1,psfig]{article}

\newcommand{\AmS}{{\protect\the\textfont2
  A\kern-.1667em\lower.5ex\hbox{M}\kern-.125emS}}

\title{Photoproduction of the Hypertriton}

\author{T. Mart$^{\rm a}$, D. Kusno\address{Jurusan Fisika, FMIPA, 
        Universitas Indonesia, Depok 16424, Indonesia},
        C. Bennhold\address{Dept. of Physics, The George Washington
        University, Washington, D.C. 20052, USA}, L. Tiator$^{\rm c}$, and
        D. Drechsel\address{Institut f\"ur Kernphysik, Johannes 
        Gutenberg-Universit\"at, 55099 Mainz, Germany}}

\begin{document}
\maketitle

\begin{abstract}
In the framework of the impulse approximation we study the photoproduction 
of the hypertriton $^3_{\Lambda}$H by using realistic $^3$He wave functions 
obtained as solutions of Faddeev equations with the Reid soft-core potential 
for different $^3_{\Lambda}$H wave functions. We obtain relatively small 
cross sections of the order of 1 nb. We also find 
that the influence of Fermi motion is important, while the effect of 
different off-shell assumptions on the cross section is not too significant.
\end{abstract}

\section{INTRODUCTION}
Photoproduction of the hypertriton can provide new information on the $YN$ 
interaction, which up to now is only poorly known from the available 
$YN$ scattering data. Being the lightest hypernucleus, the hypertriton 
is obviously the first system in which the $YN$ potential, including the 
interesting $\Lambda$-$\Sigma$ conversion potential, can be tested in the 
nuclear environment. This is also supported by the fact
that neither the $\Lambda N$ nor the $\Sigma N$ interactions are 
sufficiently strong to produce a bound two-body system. Therefore 
the hypertriton will play an important role in hypernuclear physics,
similar to the deuteron in nuclear physics. Although 
hypernuclear systems have been extensively studied for a wide range 
of nuclei by means of hadronic processes such as stopped and low momentum 
kaon induced reactions, ${\rm A}(K,\pi) _{\Lambda} {\rm B}$, as well as
${\rm A}(\pi,K) _{\Lambda} {\rm B}$ reactions, electromagnetic productions 
will, at some point, be required for a complete understanding of 
hypernuclear spectra. 

In this work we investigate the reaction 
$^3{\rm He}(\gamma,K^+)^3_{\Lambda}{\rm H}$, i.e. 
the incoming real photon interacts with a nucleon (proton) in $^3$He 
creating a lambda which combines with the other two nucleons to form 
the bound hypertriton and a positively charged kaon which exits the nucleus. 
To our knowledge, no analysis has been made and no experimental data are 
available for this reaction. The only related work is due to 
Komarov {\it et al.} \cite{komarov}, who investigated 
the complementary reaction, $p + d \rightarrow  K^+ +~^3_{\Lambda}$H, 
and estimated that at an incident proton energy $T_p$ = 1.13 - 3.0 GeV, 
the maximum differential cross section is well below 1 nb/sr.

\section{THE THREE-BODY WAVE FUNCTIONS}

In our formalism, the $^3$He wave functions are expanded in orbital momentum, 
spin, and isospin of the pair (2,3) and the spectator (1) with the notation
\begin{eqnarray}
\Psi_{^3{\rm He}}({\vec p},{\vec q}\, ) &=& \sum_{\alpha} \phi_{\alpha}(p,q) ~|
(Ll){\cal L},(S{\textstyle \frac{1}{2}}){\cal S}, {\textstyle \frac{1}{2}} M 
\rangle ~| (T{\textstyle \frac{1}{2}}) {\textstyle \frac{1}{2}} M_t \rangle ~,
\label{eq:hewf}
\end{eqnarray}
where ${\vec p}$ (${\vec q}\, $) denotes the momentum of the pair (spectator) 
and $\phi_{\alpha}(p,q)$ stands for numerical solutions of Faddeev 
equations using the realistic nucleon-nucleon potential \cite{kim}. In 
Eq.~(\ref{eq:hewf}) we have introduced $\alpha = \{ Ll{\cal L}S{\cal S}T \}$ 
to shorten the notation, where $L$, $S$, and $T$ are the total angular 
momentum, spin, and isospin of the pair (2,3), while for the spectator (1) the 
corresponding observables are labelled by $l$, $\frac{1}{2}$, 
and $\frac{1}{2}$, respectively.

For the hypertriton we choose the simple model developed in 
Ref.~\cite{congleton}, which should be reliable enough to obtain a first 
estimate for the photoproduction of the hypertriton. Using the same notation 
as in Eq.~(\ref{eq:hewf}) the wave function may be written as  
\begin{eqnarray}
\Psi_{^3_{\Lambda}{\rm H}}({\vec p},{\vec q}\, ) &=& \sum_{\alpha} 
\phi_{\alpha}(p,q)~| (Ll){\cal L},(S{\textstyle \frac{1}{2}}){\cal S}, 
{\textstyle \frac{1}{2}} M \rangle  ~,
\label{eq:hypwf}
\end{eqnarray}
where  $\phi_{\alpha}(p,q)$ is given by the two separable wave 
functions of the deuteron and the lambda, $\phi_{\alpha}(p,q) ~=~ 
\Psi^{(L)}_{d}(p)~\varphi_{\Lambda}(q)$, with the lambda part of the wave 
functions obtained by solving the Schr\"odinger equation for a particle moving 
in the $\Lambda$-$d$ effective potential.

While using this simple wave function for most calculations, we
also compared with the results for the correlated Faddeev wave function
of Ref.~\cite{miyagawa2} in order to probe the sensitivity of the cross
section to different descriptions of the hypertriton. In spite of a more 
complicated structure, the wave function may still be written in the form of 
Eq.~(\ref{eq:hewf}).

\section{THE CROSS SECTIONS}

In the lab system the cross section for kaon photoproduction off $^3$He is 
\begin{eqnarray}
  \label{eq:cssimple}
 \frac{d\sigma_{\rm T}}{d\Omega_K} &=& \frac{|{\vec q}_K^{\rm ~ c.m.}|}{|
{\vec k}^{\rm c.m.}|}~\frac{M_{\rm ^3He} 
E_{\rm ^3_{\Lambda}H}}{64\pi^2W^2}~\sum_{\epsilon}
\sum_{M,M'}~\left|T_{\rm fi}\right|^2 ~,
\end{eqnarray}
where ${\vec q}_K^{\rm ~ c.m.}$, ${\vec k}^{\rm c.m.}$, and $W$ represent
the momentum of kaon, photon, and the total energy in the c.m. system,
respectively.

Since both initial and final states of the nucleus are unpolarized, 
the sums over the spin projections can be performed by means of
${\displaystyle \sum_{M,M'}~\left|T_{\rm fi}\right|^2 = 
\sum_{\Lambda,m_{\Lambda}}~|T^{(\Lambda)}_{m_{\Lambda}}|^2 }$, 
with 
\begin{eqnarray}
T^{(\Lambda)}_{m_{\Lambda}} &=& \sqrt{\frac{2}{\pi}}~\biggl( 
\frac{E_{^{3}{\rm He}} E_{^{3}_{\Lambda}{\rm H}}}{M_{^{3}{\rm He}} 
M_{^{3}_{\Lambda}{\rm H}}} \biggr)^{\frac{1}{2}} \times \nonumber\\
&& \sum_{\alpha,{\alpha}',n}
\left[~ i^{n}~\hat{n}\hat{\cal L}\hat{S}' \hat{S}  
(-1)^{n+{\cal S}-\frac{1}{2}}
\left\{ \begin{array}{lll}  {\cal S}' & {\cal S} & n \\  
\frac{1}{2} & \frac{1}{2} & 1 \end{array} \right\}
\left\{ \begin{array}{lll}  {\cal L} & {\cal S} & \frac{1}{2} \\  
 L & {\cal S}' & \frac{1}{2}\\l&n&\Lambda \end{array} \right\}
 \delta_{LL'}\delta_{S1}\delta_{T0} ~ \times  \right. \nonumber\\
&& \int d^{3}{\vec q}~p^{2}dp~\biggl( \frac{m_{\rm i} m_{\rm f}}{
E_{\rm i} E_{\rm f}} \biggr)^{\frac{1}{2}} \varphi_{\Lambda}(q\, ')~
\Psi_{d}^{(L)}(p)~\phi_{\alpha}(p,q)~
\Bigl[{\bf Y}^{(l)}(\hat{{\vec q}\, })\otimes
{\bf K}^{(n)}\Bigr]^{(\Lambda)}_{m_{\Lambda}} \Biggr] , ~
\label{eq:tfi3}
\end{eqnarray}
where ${\bf K}^{(0)}=L$ and ${\bf K}^{(1)}={\vec K}$, i.e. the
spin-independent and spin-dependent elementary amplitudes \cite{terry}. Since 
the tensor ${\bf K}^{(n)}$ contains complicated functions of the integration 
variables ${\vec q}$ and $\hat{{\vec q}\, }=\Omega_q$, the 
integral in Eq.~(\ref{eq:tfi3}) has to be performed numerically.

The tensor operators, $[ {\bf Y}^{(l)}(\hat{{\vec q}\, }) \otimes 
{\bf K}^{(n)} ]^{(\Lambda )}_{m_{\Lambda}}$, which determine the 
specific nuclear transitions in the reaction, are given in Ref. \cite{terry}.
In contrast to the case of pion production off $^3$He \cite{tiator2}, 
the tensor operators in our case are simplified by the approximation 
that the hypertriton wave function only contains the partial wave with 
$l'=0$. However, for future studies involving all partial waves of the 
advanced hypertriton model \cite{miyagawa2}, the complete operator will be 
needed. For this purpose, we have also derived the form of Eq.~(\ref{eq:tfi3})
for the more general case \cite{terry1}.

\section{RESULTS AND DISCUSSION}

As a check of our calculations and computer codes, we compare the full 
result with two simple approximations. First, we reduce the cross
section by allowing only $S$--waves to contribute to the amplitudes
in Eq.~(\ref{eq:tfi3}). In this approximation we obtain
\begin{eqnarray}
  \label{ratio}
  \frac{d\sigma ({\rm ^3He})}{d\sigma (p)} &\approx& \frac{|{\tilde 
{{\vec K}\, }}|^2}{|{\vec K}|^2} ~\approx~ 1.8\times 10^{-3} ~,
\end{eqnarray}
where the tilde denotes the integration over the internal momenta 
${\vec p}$ and ${\vec q}$ weighted by the two wave functions. At $k=1.8$ GeV 
the elementary reaction model of Ref.~\cite{williams} yields a maximum cross 
section of about 500 nb/sr. Hence, the corresponding cross section on 
$^3$He will be about 1 nb/sr at most. 

\begin{figure}[!ht]
\begin{minipage}[t]{80mm}
\centerline{\psfig{figure=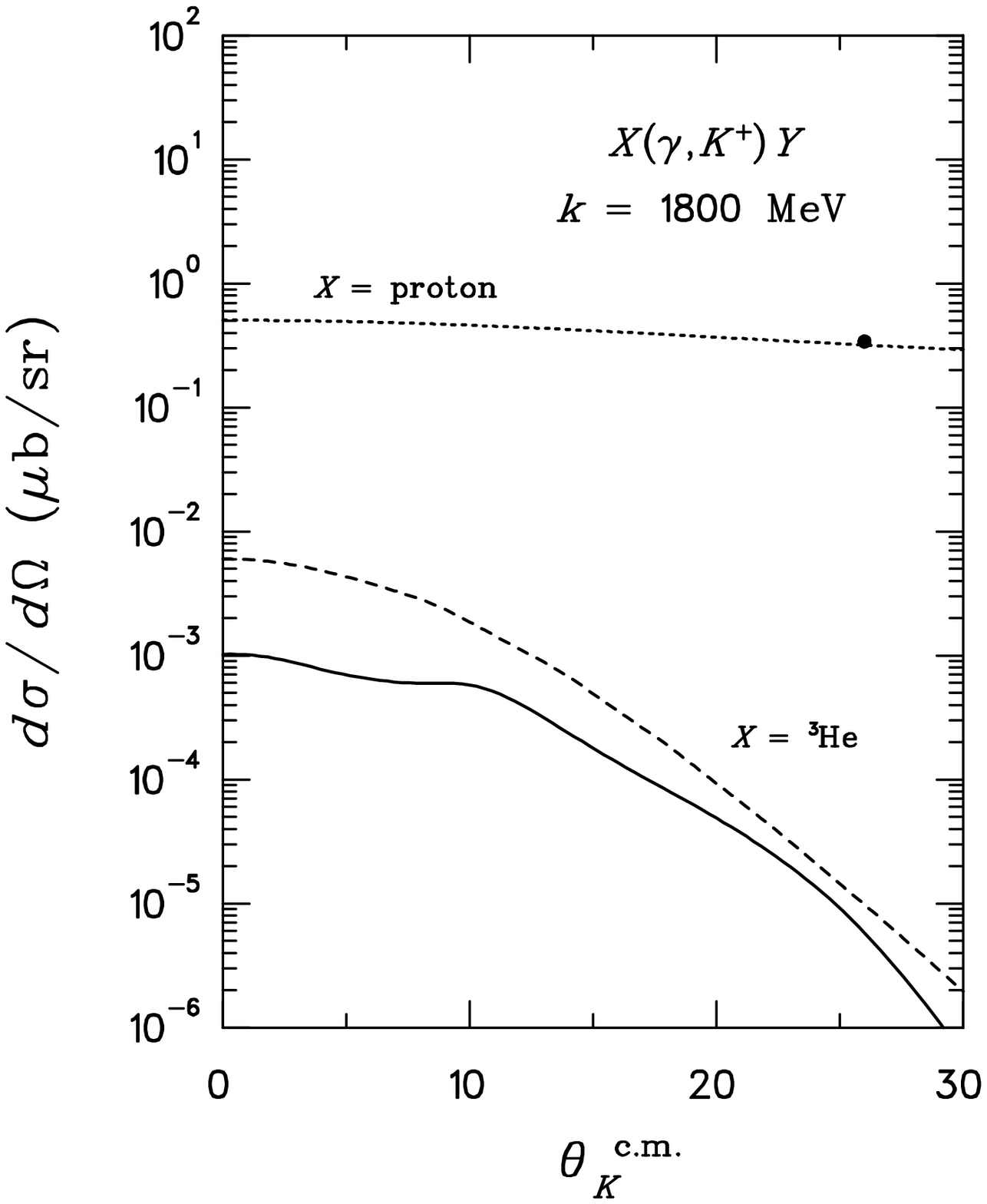,width=8cm}}
\caption{Differential cross section for kaon photoproduction off the proton
 and $^3$He as function of kaon angle. The elementary reaction (dotted 
 line) is taken from  Ref.~\protect\cite{williams} and the corresponding 
 experimental datum is from Ref.~\protect\cite{fe}. The dashed line shows the
 approximation for production off $^3$He calculated from Eq.~(\ref{msabit}), 
 the solid line represents the exact calculation using $S$-waves.}
\label{fig:ffq2}
\end{minipage}
\hspace{\fill}
\begin{minipage}[t]{75mm}
\centerline{\psfig{figure=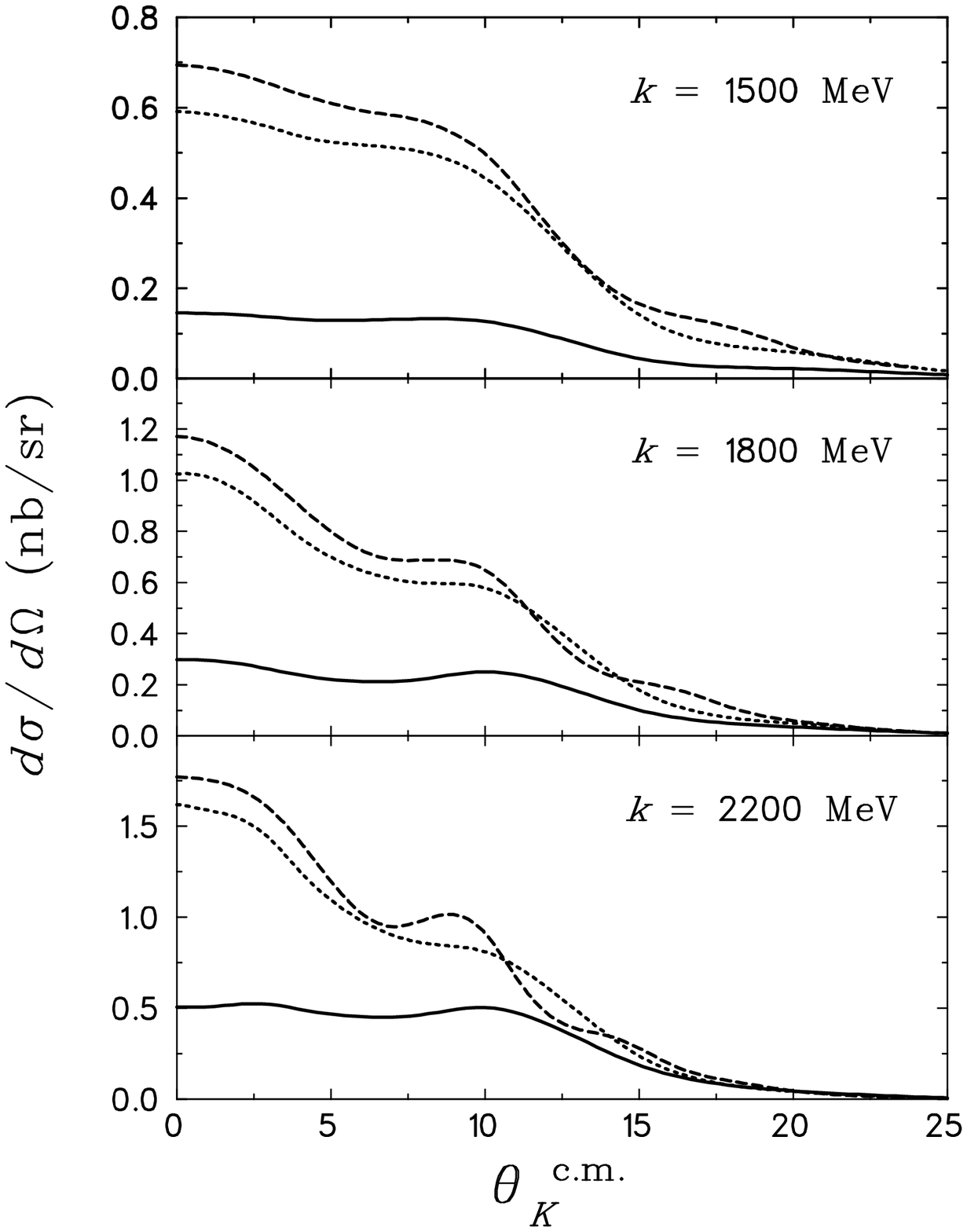,width=7.5cm}}
\caption{The cross section for kaon photoproduction off $^3$He at three
 different excitation energies. The dotted curves are obtained from the 
 the calculation with $S$--waves only and the simple hypertriton wave
function, the dashed curves are obtained with $S$--waves only
and the correlated Faddeev wave function of Ref.~\protect\cite{miyagawa2},
while the solid curves show the result after using all of the partial waves
and the simple hypertriton wave function.}
\label{fig:varwave}
\end{minipage}
\end{figure}

As a second approximation, we consider the struck nucleon inside $^3$He  
as having a fixed momentum. Therefore, the $L$ and
${\vec K}$ amplitudes can be factored out of the integral and the cross
section off $^3$He may be written in terms of the nuclear form factor 
$F(Q)$ and the elementary differential cross section,
\begin{eqnarray}
  \label{msabit}
 \frac{d\sigma_{\rm T}}{d\Omega_K} &=&{\textstyle \frac{1}{9}}~ W_A^2~|F(Q)|^2 
~\left( \frac{d\sigma_{\rm T}}{d\Omega_K}\right)_{\rm proton} ~,
\end{eqnarray}
where $W_A$ represents the kinematical factor given, i.e., in 
Ref. \cite{terry}.
The result is displayed in Fig.~\ref{fig:ffq2}. The nuclear cross section 
at forward angles is smaller than in the case of elementary kaon production by 
two orders of magnitude. As $\theta_K^{\rm c.m.}$ increases, the cross 
section drops quickly, because the momentum transfer to the nucleus increases 
as function of $\theta_K^{\rm c.m.}$. Therefore, the cross section is very 
small. The underlying reason is the lack of high momentum components in the 
$^3_\Lambda{\rm H}$ wave function. Since the momentum transfers are high, 
the lambda momentum is high as well, which inhibits hypernuclear formation.

Figure \ref{fig:ffq2} also shows the significant difference between the cross
sections calculated with the approximation of Eq.~(\ref{msabit}) and 
the full result obtained from Eq.~(\ref{eq:cssimple}). This discrepancy is  
due to the {\sl factorization} approximation. It can be seen that in 
the full calculation the integrations of both spin-independent and 
spin-dependent amplitudes over the internal momenta lead to 
destructive interferences and a further reduction of the cross section. 

In contrast to our previous conjecture, Fig.~\ref{fig:varwave} shows the 
significance of the higher partial waves which reduce the cross section by a 
factor of more than three. The reason can be traced back to 
Eq.~(\ref{eq:tfi3}). In spite of its small amplitude, the transition 
from $\alpha=8$ to $\alpha ' = 1$ may not be neglected, since $\alpha '=1$ 
is the most likely state in the hypertriton. We also note that the angular 
momentum part of the tensor amplitude in Eq.~(\ref{eq:tfi3}) yields a 
considerably large contribution for this transition. In comparison, the 
higher partial waves in pion photo- and electroproduction 
decrease the cross section by at most 15\% and 20\%, respectively.
\begin{figure}[ht]
\begin{minipage}[t]{80mm}
\centerline{\psfig{figure=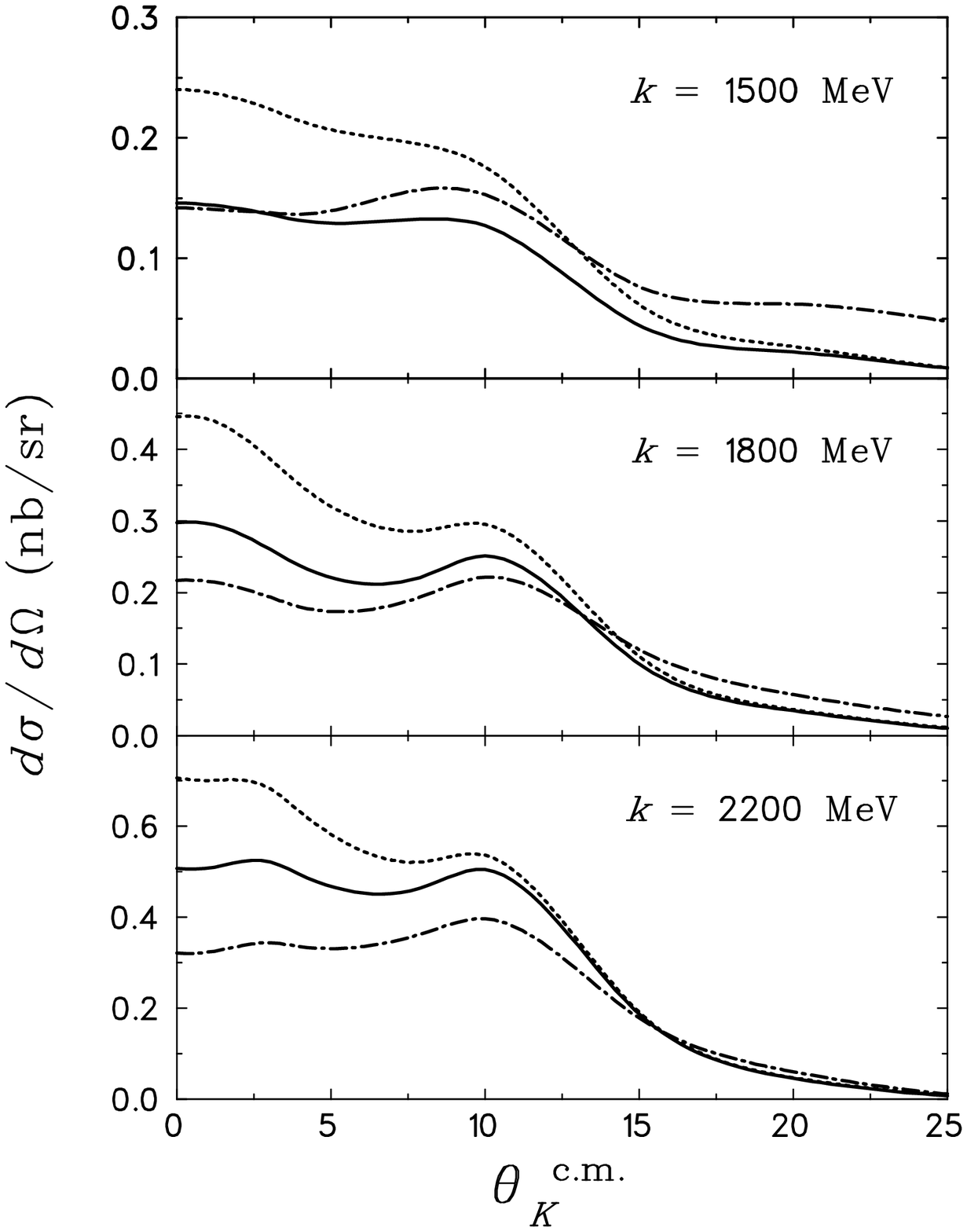,width=8cm}}
\caption{The influence of Fermi motion on the differential cross section
 at three different photon energies. The dash-dotted (dotted) curve is the 
 {\sl frozen nucleon} (average momentum) approximation, the solid
 curve shows the exact result.}
\label{fig:varfm}
\end{minipage}
\hspace{\fill}
\begin{minipage}[t]{75mm}
\centerline{\psfig{figure=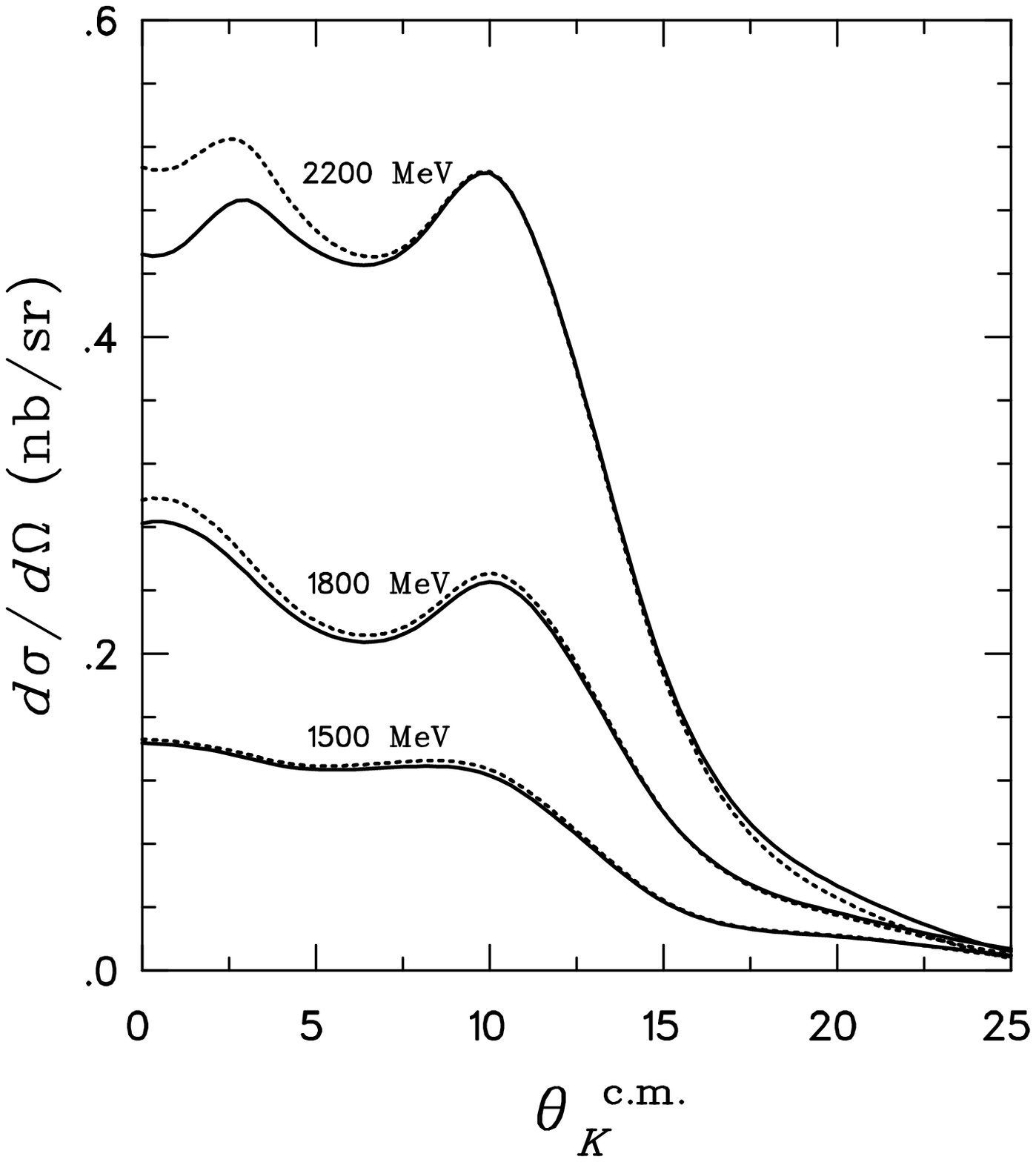,width=7.5cm}}
\caption{The effect of different off-shell assumptions on the cross section
 calculated at three different energies.
 The dotted curves have been calculated with the initial nucleon 
 on-shell, the solid curves with the final hyperon  on-shell.}
\label{fig:varos}
\end{minipage}
\end{figure}

Since the $(\gamma,K)$ process is a high momentum transfer process
and the simple analytical hypertriton wave function used until now
contains no short-range correlations, we also show in Fig.~\ref{fig:varwave}
a comparison with the correlated three-body wave function of 
Ref.~\cite{miyagawa2} that includes a proper short-range behavior. While 
the cross section obtained with the Faddeev wave function shows more 
structures, the differences are only of order 10-20$\%$. The absence of 
short-range correlations in the simple hypertriton model does only become 
obvious for much larger momentum transfers. 

The small size of the cross section obtained raises the question
of a possible significance of two-step processes, such as
$\gamma + p \rightarrow p + \pi^0 \rightarrow K^+ \Lambda$.
Two-step processes were studied in Ref.~\cite{kamalov95} for pion
photoproduction on $^3$He and found to be significant only at
$Q^2$ much larger than in our case. Ref.~\cite{fix97} also 
included  these processes in $\eta$ photoproduction on the deuteron
and found small effects.  

We have investigated the contribution of non-localities generated
by Fermi motion in the initial and final nuclei. As in former studies 
\cite{tiator2}, an exact treatment of Fermi motion is included in the 
integrations over the  wave functions in Eq.~(\ref{eq:tfi3}), while 
a local approximation can be carried out by freezing the operator 
at an average nucleon momentum 
$\langle {\vec k}_1 \rangle = -\kappa (A-1){\vec Q}/2A$, 
where $A=3$ in this case. A value of $\kappa =0$  
corresponds to the {\sl frozen nucleon} approximation, whereas  
$\kappa =1$  yields the average momentum approximation. 
The latter case has been shown to yield satisfactory results 
for pion photoproduction in the $s$- and $p$-shells \cite{tiator3}. 
Figure~\ref{fig:varfm} compares the cross sections calculated in the 
two approximations with the exact calculation. 
There appears a systematic discrepancy 
between the calculation with Fermi motion and the one with the average 
momentum approximation at all energies. Unlike in pion 
photoproduction, the average momentum approximation cannot simulate 
Fermi motion in kaon photoproduction, and the 
discrepancies between the different methods, especially near forward 
angles, are too significant to be neglected. 

Finally, we show the effect of different off-shell assumptions on the
cross section in Fig.~\ref{fig:varos}. The nucleons 
in the initial and final states are clearly off-shell. However, the 
elementary amplitudes are constructed for on-shell nucleons in the 
initial and final states. For this reason, we test the prescriptions 
given in Ref.~\cite{tiator2}, i.e. we assume that (1) the initial
nucleon is on-shell and the final hyperon off-shell, and (2) the final 
hyperon is on-shell and the initial nucleon off-shell. Both assumptions are 
compared in Fig.~\ref{fig:varos}, where we see that the difference is not too 
significant. The largest discrepancy of 10\% occurs at $k=2200$ MeV 
in the forward direction as was also observed in the case 
of pion photoproduction. 

In conclusion, we find that the hypertriton photoproduction could provide 
a sensitive test of the hypertriton wave functions. The small cross sections 
and the required high resolution measurements would become a new challenge 
to experimentalists.

\vspace{7mm}

\noindent {\bf ACKNOWLEDGMENTS}

\vspace{3mm}

The authors thank Dr. K. Miyagawa for providing the advanced model of the 
hypertriton wave function.
The work of TM and DK was supported in part by the University Research for 
Graduate Education (URGE) grant. CB has been supported in part by US 
Department of Energy under contract no. DE-FG02-95-ER40907.


\begin{thebibliography}{19}
\bibitem{komarov} V. I. Komarov, A. V. Lado, and Yu. N. Uzikov,
                  J. Phys. G 21 (1995) L69.
\bibitem{kim} R. A. Brandenburg, Y. E. Kim, and A. Tubis, Phys. Rev. C
              12 (1975) 1368.
\bibitem{congleton} J. G. Congleton, J. Phys. G 18 (1992) 339.
\bibitem{miyagawa2} K. Miyagawa, H. Kamada, W. Gl\"ockle, and V. Stoks,
                  Phys. Rev. C 51 (1995) 2905.
\bibitem{terry} T. Mart, L. Tiator, D. Drechsel, C. Bennhold, and K. Miyagawa,
                submitted to Nucl. Phys. A.
\bibitem{tiator2} L. Tiator, A. K. Rej, and D. Drechsel, Nucl. Phys.
                  A 333 (1980) 343.
\bibitem{terry1} T. Mart, Ph.D. Thesis, Universit\"at Mainz, 1996 (unpublished).
\bibitem{williams} R. A. Williams, C.-R. Ji, and S. R. Cotanch, Phys. Rev. 
                   C 46 (1992) 1617.
\bibitem{fe} P. Feller {\it et al}., Nucl. Phys. B 39 (1972) 413.
\bibitem{kamalov95} S. S. Kamalov, L. Tiator, and C. Bennhold,
        Phys. Rev. Lett. 75 (1995) 1288.
\bibitem{fix97} A. I. Fix and V. A. Tryasuchev, Yad.Fiz. 60 (1997) 41, 
        translated in: Phys. of At. Nucl. 60 (1997) 35.
\bibitem{tiator3} L. Tiator and L. E. Wright, Phys. Rev. C 30 (1984) 989.
\end{thebibliography}
\end{document}